\documentclass{moriond}

\newcommand{\Photo}{\includegraphics[height=35mm]{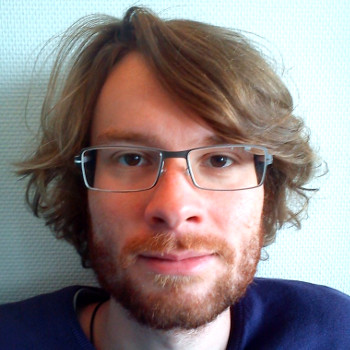}}

\begin{document}
\vspace*{4cm}
\title{PARTON ENERGY LOSS AND PARTICLE PRODUCTION AT HIGH MOMENTA FROM ALICE }

\author{R. A. BERTENS for the ALICE collaboration}

\address{Institute for Subatomic Physics, Utrecht University \\ Princetonplein 5, 3508 TA Utrecht, The Netherlands}

\maketitle\abstracts{Partons, produced in the early stages of heavy-ion collisions, lose energy while propagating through the collision medium. This energy loss can be studied by comparing particle yields in different systems (pp, p--Pb, Pb--Pb). In addition, particle yields in different systems can be used to study hadronization mechanisms. } 

\section{Introduction and analysis}
The heavy-ion program at ALICE is aimed at studying strongly interacting matter in ultra-relativistic nuclear collisions where the formation of a Quark-Gluon Plasma (QGP), a deconfined state of quarks and gluons, is expected \cite{qgp}. Hard partons that propagate through this matter are predicted to lose energy via (multiple) scattering and gluon radiation. As a result, {\ensuremath{p_{\mathrm{T}}}} spectra of final state hadrons and jets will be modified with respect to yields derived from a simple superposition of incoherent proton-proton collisions. This modification, quantified by the \emph{nuclear modification factor}, {\ensuremath{R_\mathrm{AA}}}, is used to study parton energy-loss mechanisms and medium properties. Disentangling energy-loss signatures from initial state nuclear effects which may also modify transverse momentum spectra - such as nuclear PDF shadowing \cite{nPDF} - requires a comparison of the {\ensuremath{R_\mathrm{AA}}} to the nuclear modification factor for proton-nucleus collisions, called {\ensuremath{R_\mathrm{pA}}}. These proceedings give an overview of recent ALICE results on the nuclear modification factor for Pb--Pb collisions as well as p--Pb collisions. 

ALICE \cite{alicedetector} is a general-purpose heavy-ion experiment at CERN. Its central barrel includes an Inner Tracking System (ITS), Time Projection Chamber (TPC), Transition Radiation (TRD) and Time Of Flight (TOF) detector used for tracking (ITS, TPC) and identification (TPC, TOF, TRD) of charged particles. At high momenta ($>$ 1 {\ensuremath{\mathrm{GeV}\kern-0.05em/\kern-0.02em c}}) identification is complemented by a small acceptance ring imaging Cherenkov detector. Neutral mesons are reconstructed using an electromagnetic calorimeter; muons with a forward muon spectrometer.

The nuclear modification factor {\ensuremath{R_\mathrm{AA}}} is defined as 
\begin{equation}\label{eq:raa}
    {\ensuremath{R_\mathrm{AA}}} = \frac{\mathrm{d}^2 {\ensuremath{N_\mathrm{AA}}} / \mathrm{d} {\ensuremath{p_{\mathrm{T}}}} \mathrm{d}\eta}
    {\langle {\ensuremath{T_\mathrm{AA}}} \rangle \cdotp \mathrm{d}^2 {\ensuremath{\sigma_\mathrm{pp}}} / \mathrm{d} {\ensuremath{p_{\mathrm{T}}}} \mathrm{d}\eta } \nonumber
\end{equation}
where $\mathrm{d}^2 {\ensuremath{N_\mathrm{AA}}} / \mathrm{d} {\ensuremath{p_{\mathrm{T}}}} \mathrm{d}\eta$ represents the differential particle yield in nucleus-nucleus collisions and $\mathrm{d}^2 {\ensuremath{\sigma_\mathrm{pp}}} / \mathrm{d} {\ensuremath{p_{\mathrm{T}}}} \mathrm{d}\eta$ is the cross-section in proton-proton collisions. The nuclear overlap function $\langle {\ensuremath{T_\mathrm{AA}}} \rangle$ is derived from a Glauber model \cite{glauber} and proportional, in each centrality class, to the number of binary collisions $\langle {\ensuremath{N_{\mathrm{coll}}}} \rangle$. At high {\ensuremath{p_{\mathrm{T}}}} and in the absence of medium effects the {\ensuremath{R_\mathrm{AA}}} is expected to be 1; at low momenta, the spectral shape is dominated by soft processes and such a scaling is not expected to hold \cite{charged_raa}. As QGP formation is not predicted in pA collisions, {\ensuremath{R_\mathrm{pA}}} (measured similarly) can be used to disentangle (cold) nuclear effects from QGP effects. 

\section{{\ensuremath{R_\mathrm{AA}}} and {\ensuremath{R_\mathrm{pA}}} of (identified) particles and jets}
\begin{figure}
\begin{minipage}{0.4\linewidth}
\centerline{\includegraphics[height=5.1cm]{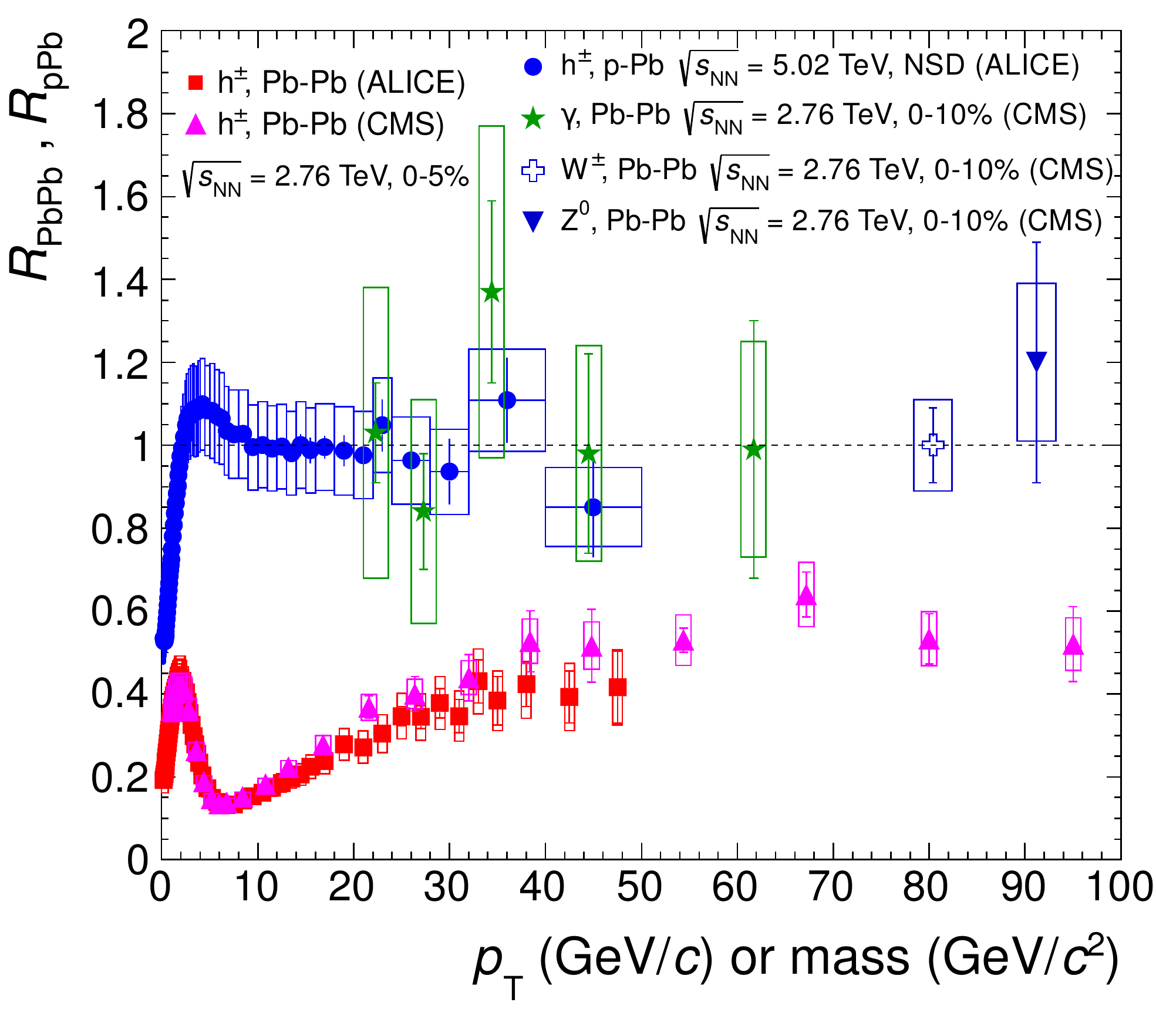}}
\caption[]{Chared hadron {\ensuremath{R_\mathrm{AA}}} in central collisions and {\ensuremath{R_\mathrm{pA}}} at mid-rapidity~\cite{charged_rpb}.}
\label{fig:raa_general_1}
\end{minipage}
\hspace{.5cm}
\begin{minipage}{0.55\linewidth}
\centerline{\includegraphics[height=5.1cm]{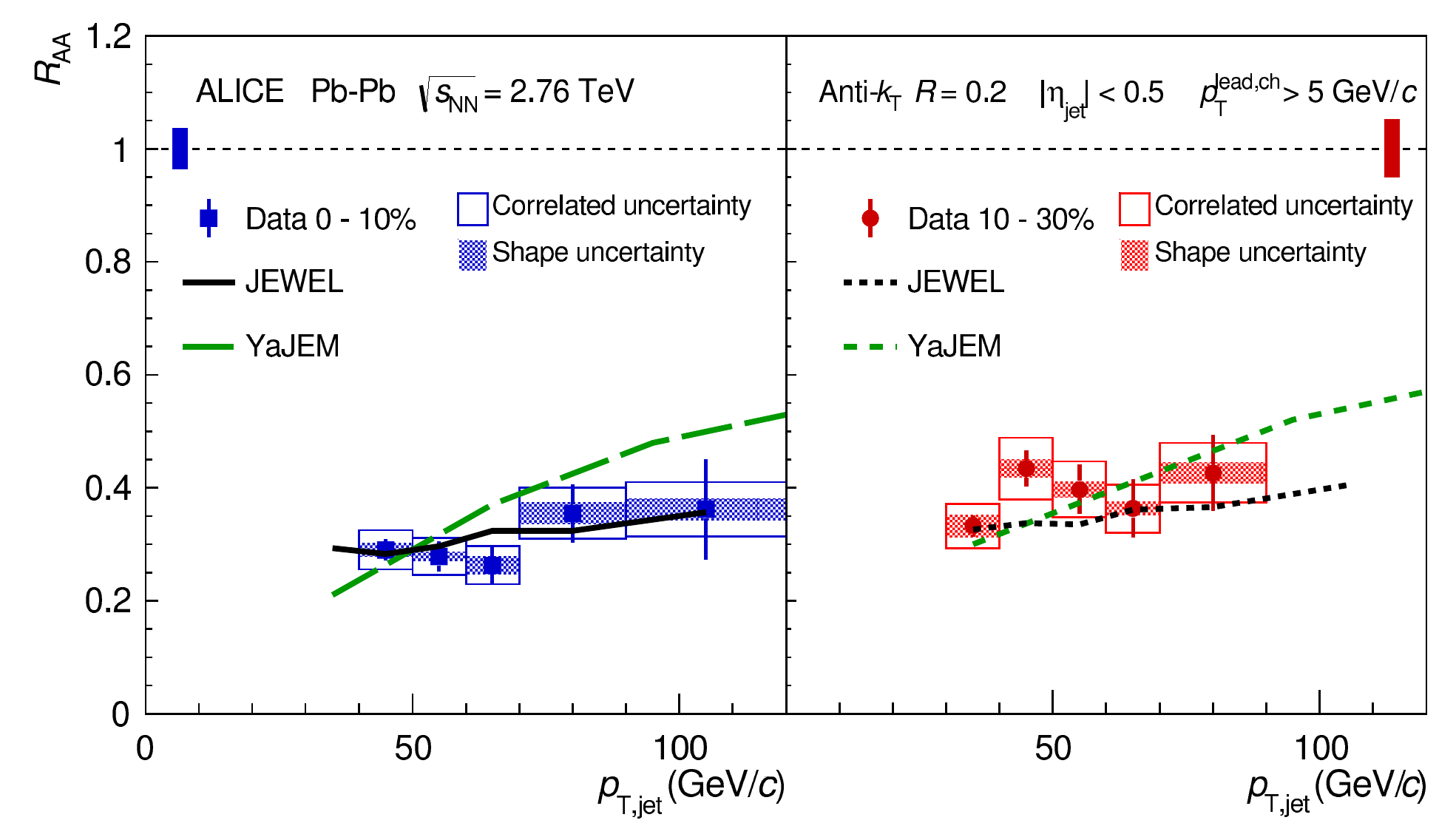}}
\caption[]{{\ensuremath{R_\mathrm{AA}}} of $R=0.2$ full jets in central and mid-peripheral collisions compared energy-loss models~\cite{jets_raa}.}
\label{fig:full_jets_1}
\end{minipage}

\end{figure}
\begin{figure}
\begin{minipage}{0.35\linewidth}
    \centerline{\includegraphics[height=4.8cm]{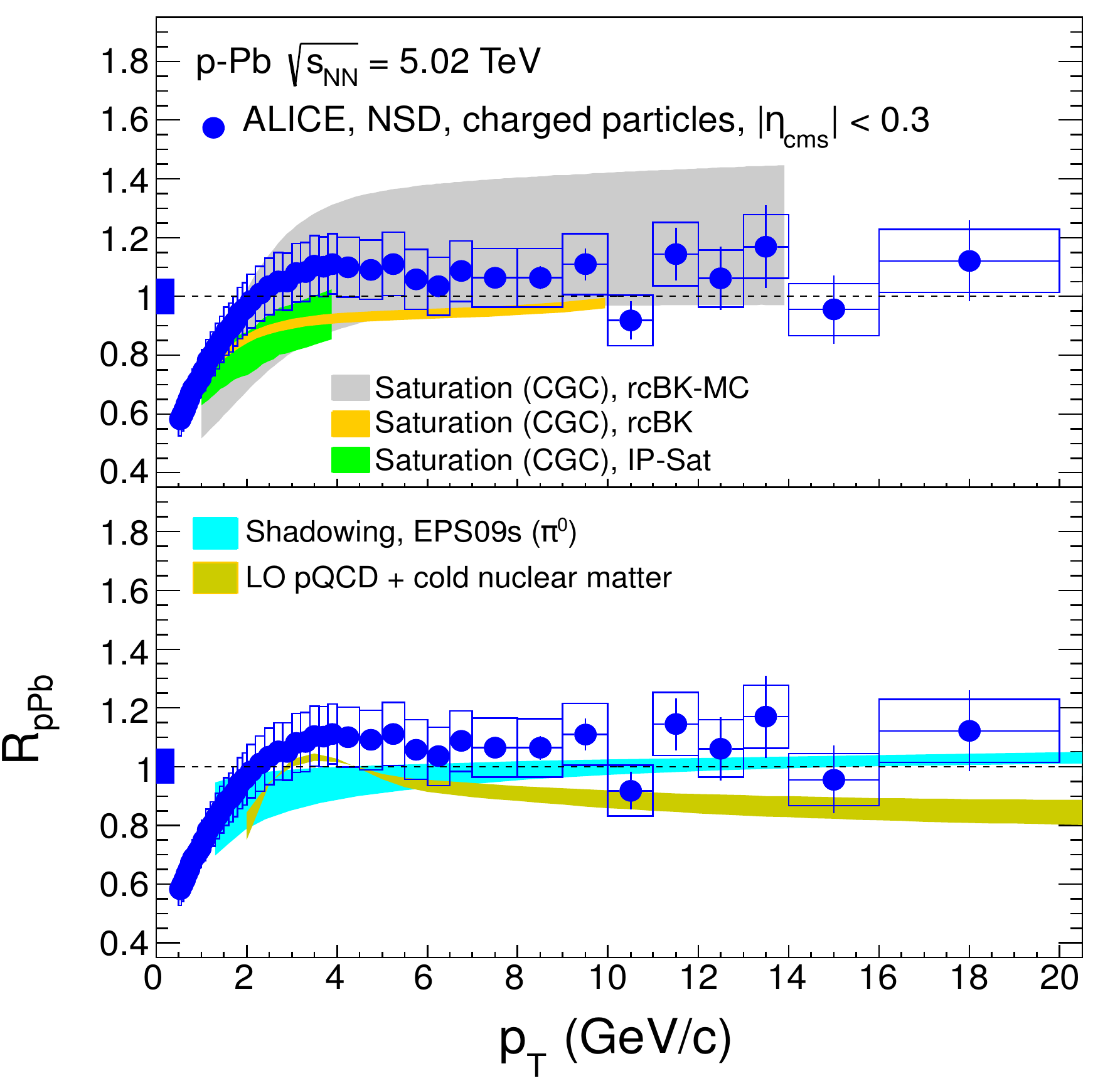}}
\caption[]{Charged hadron {\ensuremath{R_\mathrm{pA}}} compared to different models~\cite{charged_rpb}.}
\label{fig:raa_general_2}
\end{minipage}
\hspace{.5cm}
\begin{minipage}{0.6\linewidth}
    \centerline{\includegraphics[height=5cm]{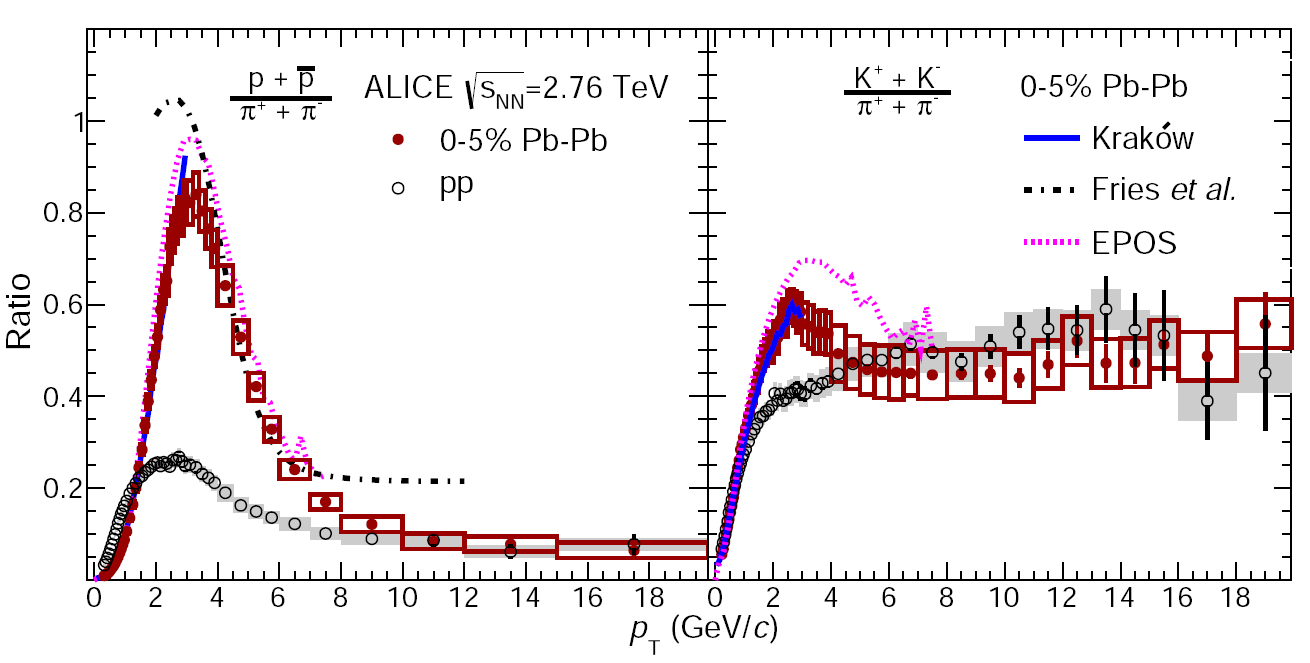}}
\caption[]{Particle ratios in pp and central Pb--Pb collisions~\cite{identified_ratios}. Theoretical predictions refer to Pb--Pb collisions.}
\label{fig:identified_hadrons_1}
\end{minipage}
\end{figure}

Fig. \ref{fig:raa_general_1} shows the charged particle {\ensuremath{R_\mathrm{AA}}} measured at {\ensuremath{\sqrt{s_{\mathrm{NN}}}}} = 2.76 TeV in central collisions compared to the charged hadron {\ensuremath{R_\mathrm{pA}}} at {\ensuremath{\sqrt{s_{\mathrm{NN}}}}} = 5.02 TeV and the {\ensuremath{R_\mathrm{AA}}} of particles which are \emph{not} sensitive to QCD dynamics ({\ensuremath{\gamma}}, W$^{\pm}$, {Z$^{0}$}). {\ensuremath{R_\mathrm{AA}}} of the {\ensuremath{\gamma}}, W$^{\pm}$ and {Z$^{0}$} is 1 within uncertainties, confirming the $\langle {\ensuremath{N_{\mathrm{coll}}}} \rangle$ scaling. The suppression of the charged hadron yield ({\ensuremath{R_\mathrm{AA}}} $<$ 1) in Pb--Pb collisions is not seen in p--Pb collisions ({\ensuremath{R_\mathrm{pA}}} = 1), which indicates that the suppression in Pb--Pb collisions is a result of final state effects, most likely parton energy loss. Similar behavior is observed in the {\ensuremath{R_\mathrm{AA}}} of jets, shown in Fig. \ref{fig:full_jets_1}. The suppression of the jet yield indicates strong out-of-cone radiation of jet energy for central and semi-central collisions. Comparisons to jet energy-loss models JEWEL~\cite{jewel} and YaJEM \cite{yajem} show a qualitative agreement ($\chi^2$ of 0.368 and 1.690 respectively \cite{jets_raa}) with the data. Models based on gluon saturation \cite{cgc1,cgc2} (Fig \ref{fig:raa_general_2}, top panel) and nPDF shadowing \cite{epos1,epos2} (lower panel) predict small initial state nuclear effects at mid-rapidy in p--Pb collisions; this is confirmed by data as the measured {\ensuremath{R_\mathrm{pA}}} is in agreement with unity for {\ensuremath{p_{\mathrm{T}}}} $>$ 4 {\ensuremath{\mathrm{GeV}\kern-0.05em/\kern-0.02em c}}.

The {\ensuremath{R_\mathrm{AA}}} and {\ensuremath{p_{\mathrm{T}}}} spectra of identified particles can be used to study hadronization mechanisms. Fig. \ref{fig:identified_hadrons_1} show the ratio of proton to pion spectra and kaon to pion spectra in Pb--Pb and pp collisions. For {\ensuremath{p_{\mathrm{T}}}} $<$ 5 {\ensuremath{\mathrm{GeV}\kern-0.05em/\kern-0.02em c}} the Pb--Pb ratios are strongly enhanced with respect to the pp measurement. This enhancement is consistent with a common velocity boost~\cite{identified_ratios} (radial flow) which leads to a mass-dependent modification of the {\ensuremath{p_{\mathrm{T}}}} spectra. The Krak\'ow and EPOS \cite{krakow,epos} models, based on a hydrodynamic collision medium, are in better agreement with the data than the Fries \cite{fries} model, which assumes recombination as the dominant hadronization mechanism. In central collisions, the $\phi$-meson spectrum (not shown, see \cite{phi}) is similar in shape to the proton spectrum, supporting the dominance of radial flow as the $\phi$ is a meson with a mass close to the proton mass. At high momenta ($>$ 10 {\ensuremath{\mathrm{GeV}\kern-0.05em/\kern-0.02em c}}) the particle ratios in pp and Pb--Pb collisions are equal, indicating vacuum-like hadronization through fragmentation. 

\begin{figure}
\begin{minipage}{0.45\linewidth}
\centerline{\includegraphics[height=5.1cm]{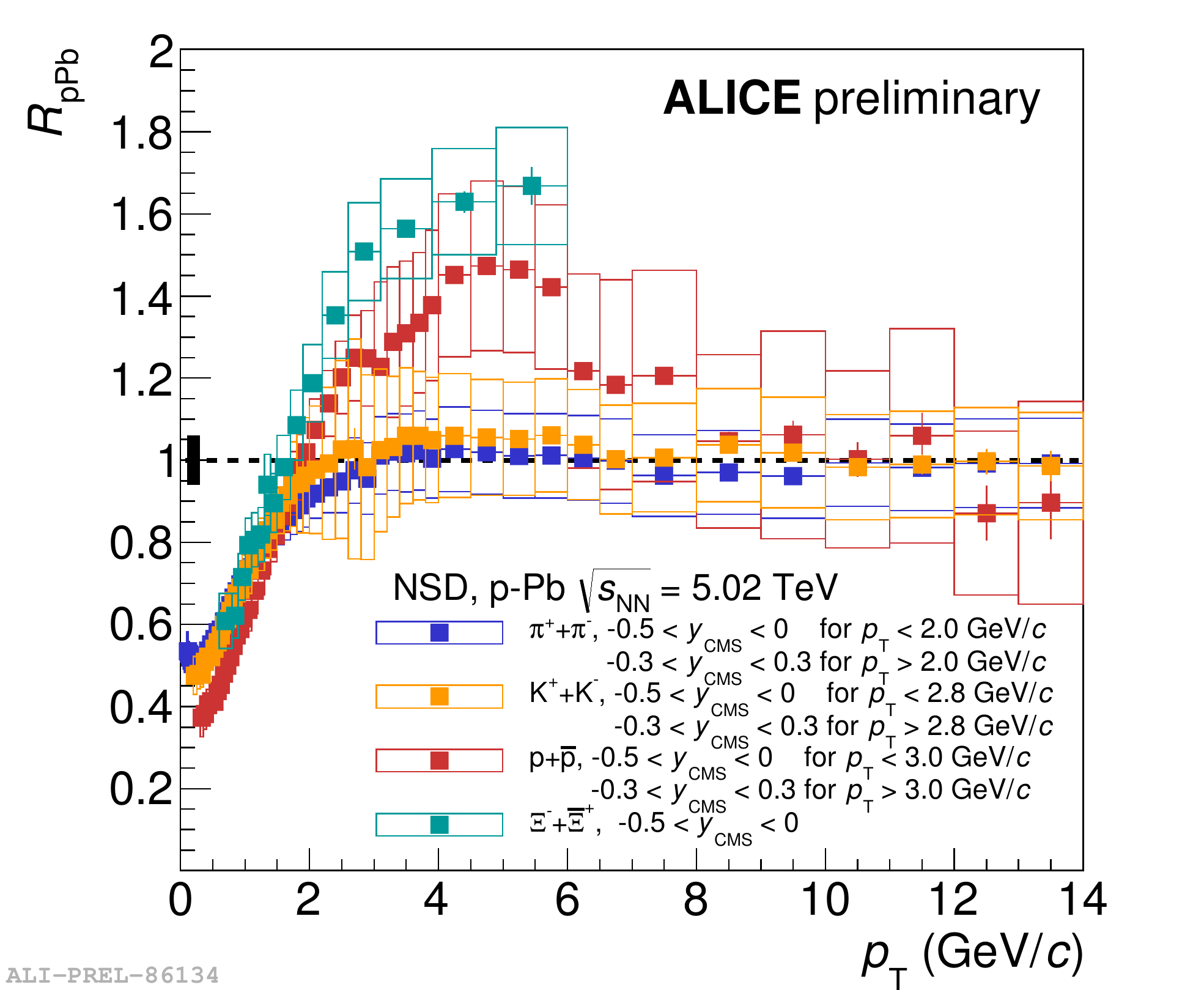}}
\caption[]{{\ensuremath{R_\mathrm{pA}}} of identified hadrons at mid-rapidity.}
\label{fig:identified_hadrons_2}
\end{minipage}
\hspace{.5cm}
\begin{minipage}{0.45\linewidth}
    \centerline{\includegraphics[height=5.1cm]{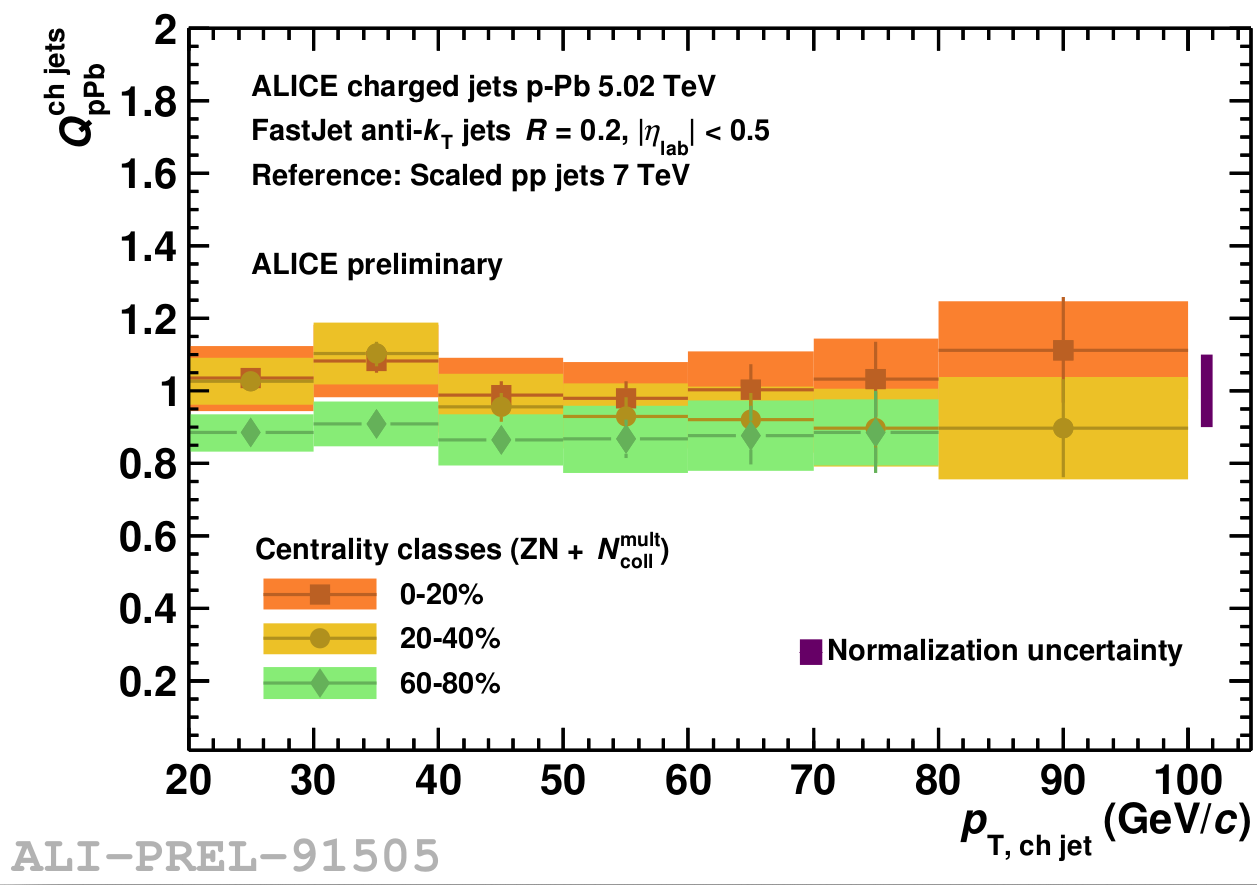}}
\caption[]{{\ensuremath{Q_\mathrm{pA}}} of charged $R=0.2$ jets with the Zero Degree Calorimeter centrality estimate.}
\label{fig:full_jets_2}
\end{minipage}
\end{figure}

Fig. \ref{fig:identified_hadrons_2} shows the {\ensuremath{R_\mathrm{pA}}} of identified hadrons ($\pi^{\pm}$, {\ensuremath{{\rm K}^{\pm}}}, {\ensuremath{\rm p\overline{p}}} and {\ensuremath{\rm \Xi\overline{\Xi}}}, the latter reconstructed in the $\Xi \rightarrow \Lambda + \pi$ channel). For {\ensuremath{p_{\mathrm{T}}}} $>$ 10 {\ensuremath{\mathrm{GeV}\kern-0.05em/\kern-0.02em c}} the {\ensuremath{R_\mathrm{pA}}} is consistent with unity (and therefore no final state effects); at intermediate momenta however, a mass ordering similar to that in Pb--Pb collisions is observed, prompting the question of whether or not this ordering is the result of collective behavior in small systems.

\section{Centrality dependence in small systems: {\ensuremath{Q_\mathrm{pPb}}}}
\begin{figure}
    \begin{minipage}{.4\linewidth}
        \centerline{\includegraphics[height=5cm]{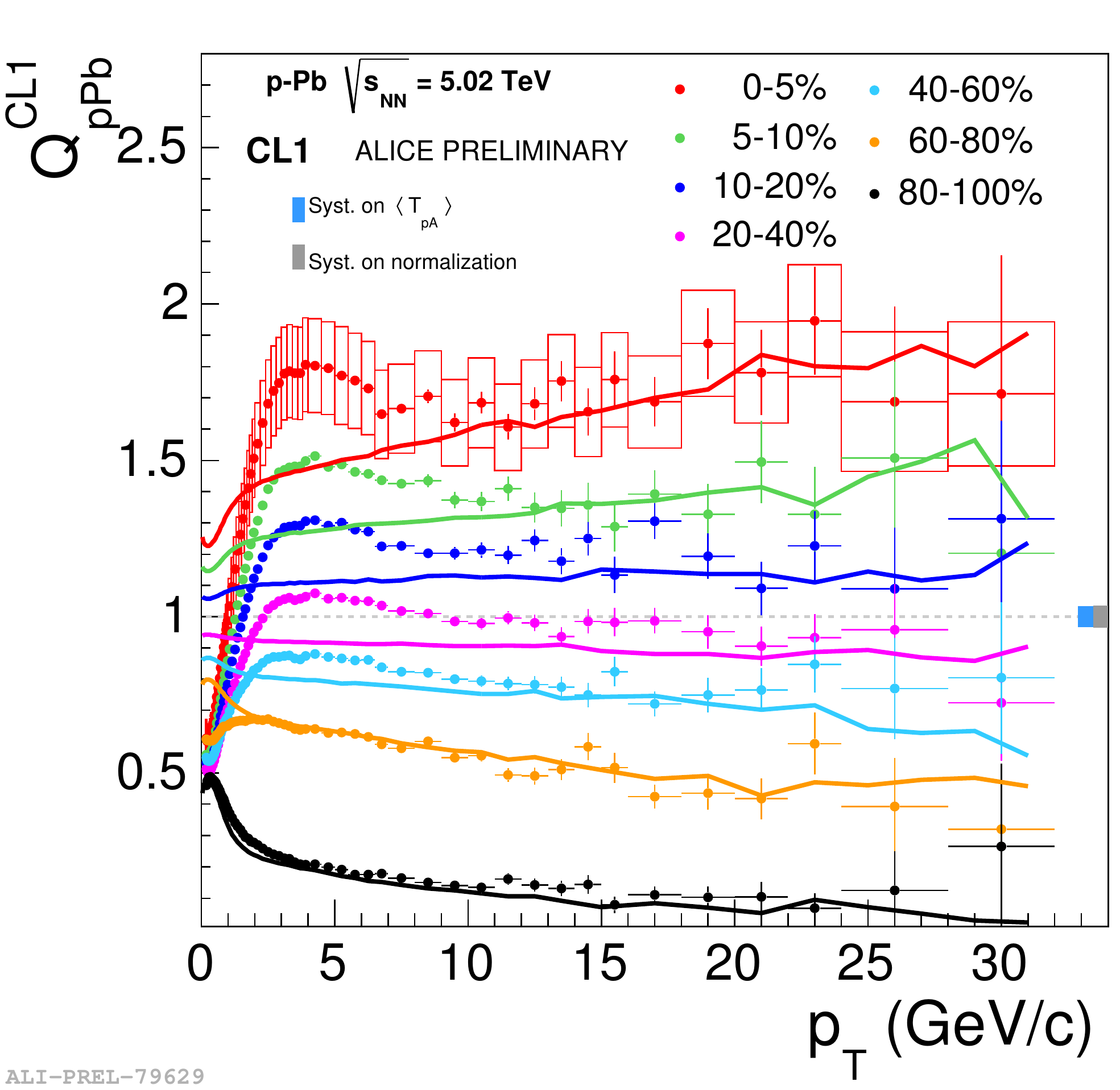}}
\caption[]{{\ensuremath{Q_\mathrm{pPb}}} with multiplicity at mid-rapidity as a centrality estimator~\cite{qpb}.}
\label{fig:QpPb_1}
    \end{minipage}
    \hspace{.5cm}
\begin{minipage}{.55\linewidth}
    \centerline{\includegraphics[height=5cm]{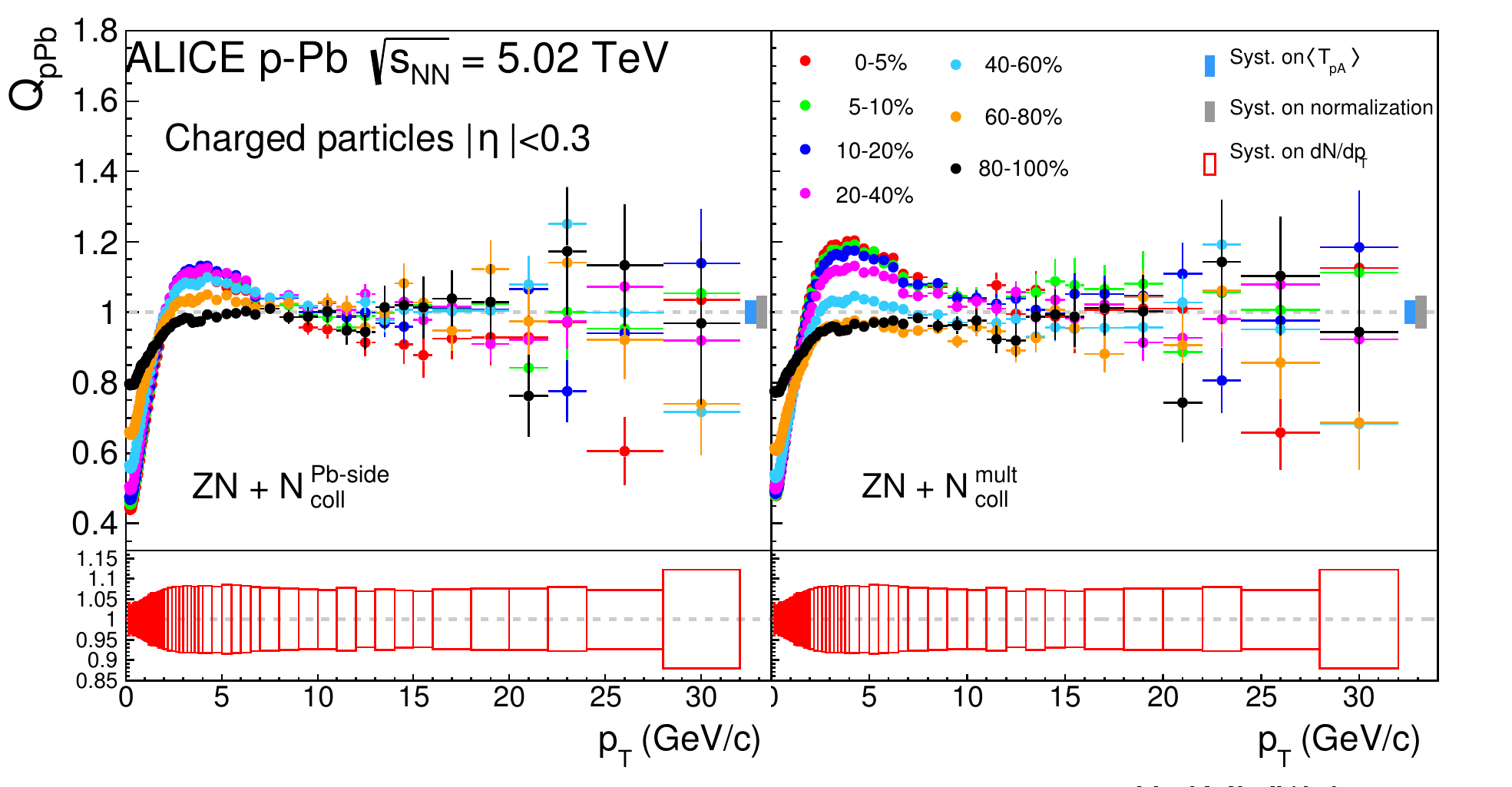}}
\caption[]{{\ensuremath{Q_\mathrm{pPb}}} with Zero Degree Calorimeter as centrality estimator for two $\langle {\ensuremath{N_{\mathrm{coll}}}} \rangle$ estimates~\cite{qpb}.}
\label{fig:QpPb_2}
\end{minipage}
\end{figure}

To characterize dynamics of small systems more precisely, a `centrality'-dependent nuclear modification factor,  {\ensuremath{Q_\mathrm{pPb}}}, is introduced in p--Pb collisions. Multiplicity fluctuates strongly for a given impact parameter~\cite{qpb}, leading to a biased {\ensuremath{Q_\mathrm{pPb}}} when centrality and $\langle {\ensuremath{N_{\mathrm{coll}}}} \rangle$ are derived directly from the number of tracks in the same $\eta$ range. This is illustrated in Fig. \ref{fig:QpPb_1}, where such a measurement (points) is compared to a model (lines) comprising incoherent PYTHIA \cite{pythia} events coupled to Glauber geometry. The observed agreement shows that the centrality dependence of the {\ensuremath{Q_\mathrm{pPb}}} is an artifact of multiplicity fluctuations and \emph{not} a result of nuclear effects.

Separating the centrality determination and the estimate of the $\langle {\ensuremath{N_{\mathrm{coll}}}} \rangle$ in $\eta$ is expected to suppress the bias from multiplicity fluctuations. Fig. \ref{fig:QpPb_2} shows the {\ensuremath{Q_\mathrm{pPb}}} measured by estimating the centrality via Zero Degree Calorimeters (situated 116 $m$ from the interaction point) and deriving $\langle {\ensuremath{N_{\mathrm{coll}}}} \rangle$ from the charged particle multiplicity at mid(left)- or forward(right) rapidities using VZERO scintillators \cite{alicedetector}. In both figures, {\ensuremath{Q_\mathrm{pPb}}} shows no centrality dependence and is in agreement with unity above {\ensuremath{p_{\mathrm{T}}}} $\approx$ 10 {\ensuremath{\mathrm{GeV}\kern-0.05em/\kern-0.02em c}}. The same is seen in Fig. \ref{fig:full_jets_2} where {\ensuremath{Q_\mathrm{pPb}}} of jets is shown - the agreement with unity at high {\ensuremath{p_{\mathrm{T}}}} in all centrality classes confirms that the jet suppression seen in Pb--Pb collisions is a medium effect.

\section{Conclusion}
The nuclear modification factor of hadrons and jets is measured in Pb--Pb and p--Pb collisions. A strong suppression is observed in Pb--Pb, but not in p--Pb measurements, confirming that partons lose energy in the medium that is formed in the collision. The {\ensuremath{R_\mathrm{AA}}} shows that relative energy loss decreases with increasing parton momenta.

From ratios of identified particle spectra it is concluded that mass rather than recombination determines the shape of spectra at low {\ensuremath{p_{\mathrm{T}}}}, whereas at higher {\ensuremath{p_{\mathrm{T}}}} fragmentation is likely to be the dominant hadronization mechanism. The {\ensuremath{R_\mathrm{pA}}} of identified hadrons exhibits a mass ordering similar to the one observed in Pb-Pb, raising interesting questions about the observation of collective effects in small systems.

\end{document}